\documentclass[aoas,MSNbibl,nameyear,seceqn,dvips]{arximspdf}
\usepackage{dcolumn}
\usepackage{graphicx}

\doi{10.1214/12-AOAS610} 
\volume{7}
\issue{2}
\pubyear{2013}
\firstpage{936}
\lastpage{954}
\setattribute{copyright}{owner}{\textup{In the Public Domain}}

\makeatletter

\newcolumntype{d}[1]{D{.}{.}{#1}}

\newcommand{\cal}{\mathcal}

\makeatother

\begin{document}
\begin{frontmatter}

\title{Spatially explicit models for inference about density in
unmarked or partially marked~populations}
\runtitle{Spatial models for unmarked populations}

\begin{aug}
\author[A]{\fnms{Richard B.} \snm{Chandler}\corref{}\thanksref{t1}\ead[label=e1]{rchandler@usgs.gov}}
\and
\author[A]{\fnms{J. Andrew} \snm{Royle}\ead[label=e2]{aroyle@usgs.gov}}
\runauthor{R. B. Chandler and J. A. Royle}
\affiliation{USGS Patuxent Wildlife Research Center}
\address[A]{USGS Patuxent Wildlife Research Center\\
12100 Beech Forest Road \\
Laurel, Maryland 20708-4039 \\
USA\\
\printead{e1}\\
\hphantom{E-mail: }\printead*{e2}} 
\end{aug}

\thankstext{t1}{Supported by the North American Breeding Bird
Survey Program.}

\received{\smonth{12} \syear{2011}}
\revised{\smonth{10} \syear{2012}}

%
\begin{abstract}
Recently developed spatial capture--recapture (SCR) models represent
a major advance over traditional capture--recapture (CR) models
because they yield explicit estimates of animal density %
instead of population size within an unknown area. Furthermore,
unlike nonspatial CR methods, SCR models account for heterogeneity
in capture probability arising from the juxtaposition of animal
activity centers and sample locations. Although the utility of SCR
methods is gaining recognition, the requirement that all individuals
can be uniquely identified excludes their use in many contexts. In
this paper, we develop models for situations in which individual
recognition is not possible, thereby allowing SCR concepts to be
applied in studies of unmarked or partially marked populations. The
data required for our model are spatially referenced counts made on
one or more sample occasions at a collection of closely spaced
sample units such that individuals can be encountered at multiple
locations. Our approach includes a spatial point process for the animal activity
centers and uses the spatial correlation in counts as information
about the number and location of the activity centers.
Camera-traps, hair snares, track plates, sound recordings,
and even point counts can yield spatially correlated count data, and
thus our model is widely applicable. A simulation study demonstrated that
while the posterior mean exhibits frequentist bias on the order of
5--10\% in small samples,
the posterior mode is an accurate point estimator as long as
adequate spatial correlation is present. Marking a subset
of the population substantially increases posterior
precision and is recommended whenever possible. We applied our model to
avian point count data collected on an unmarked population of
the northern parula (\textit{Parula americana}) and obtained a density
estimate (posterior mode) of 0.38 (95\% CI: 0.19--1.64) birds/ha. Our paper
challenges sampling and analytical conventions in ecology by
demonstrating that
neither spatial independence nor individual recognition is needed to
estimate population density---rather, spatial dependence
can be informative about individual distribution and density.
\end{abstract}

%

\end{frontmatter}

\section{Introduction}
\label{sintro}

Estimates of population density are required in basic and applied
ecological research, but are difficult to obtain for many species,
including some of the most critically endangered. A primary obstacle
faced when estimating population density is that the number of
individuals captured or detected is an unknown fraction of the actual
number present, $N$. Traditional capture--recapture (CR) methods
[\citet{seber1973}] yield estimates of $N$; however, the effective area
sampled is typically unknown, and thus density cannot be
explicitly estimated [\citet{dice38methods,wilsonAnderson85}].
This is
a well-known deficiency of traditional CR methods that makes it
difficult to interpret differences in abundance among sampling
locations and hence test hypotheses regarding spatial variation in
abundance.

An additional limitation of nonspatial CR methods is that,
even if effective sample area is known, estimators of $N$ can be biased
by unmodeled
heterogeneity in capture probability resulting from the distance
between animal ``activity centers'' and survey locations. The
definition of an activity center will depend upon the biology of the
species, but often it can be regarded as the centroid of an animal's
home range or, more generally, the first spatial moment of an animal's
locations during some time interval. 
Intuitively, individuals with activity centers close to a trap are
more likely to be captured than individuals whose activity centers are
further away. Spatial capture--recapture (SCR)
models [\citet
{efford04,borchersEfford08scr,royleYoung08,royleEA09cameraEcol}]
produce direct
estimates of density or population size for explicit spatial regions
by asserting a spatial point process model for the activity centers, and
modeling capture probability as a function of the
distance between the survey locations and the activity centers.
Although the activity centers cannot be directly
observed, information about their locations comes from the
spatial coordinates of the traps where individuals are
captured---data which have always been available but were rarely
utilized until recently.

Because SCR models overcome the limitations of CR methods without
requiring additional data, they represent a major advance in efforts
to estimate population density, and their use is becoming widespread
[\citet
{dawsonefford2009,effordetal2009,gardnerEA09DNA,sollmannetal2011,gopalaswamyetal2012}].
However, use of such methods requires that all individuals are uniquely
identifiable, which can be difficult to achieve in practice. In some
cases, 
it is not even possible to identify individuals, such as in avian point
count surveys, which involve counting unmarked individuals from
multiple points within a study area. In other cases, even when resources
are available to obtain individual recognition, the identity of many
individuals often remains unknown. For example, in camera trapping
studies [\citet{oconnell10book}],
the resulting photographs are not always sufficient
for identification due to similar markings among animals. For some
species, no natural markings are present
to aid recognition (e.g., fisher \textit{Martes pennanti},
coyote \textit{Canis latrans}), and
physically capturing individuals may be too difficult or intrusive.

In this paper, we present a model allowing for inference about density
and population size when individuals cannot be uniquely
identified nor detected with certainty. Our model requires 
spatially correlated count data from
sample locations in close proximity to one another such that
individuals can be detected at multiple locations.
Rather than viewing the spatial correlation as an inferential
obstacle, we utilize the correlation as information about distribution
and population size.
We develop our model by regarding the encounter frequencies for each
individual, or for a subset of individuals, as latent or missing data,
and we provide a Bayesian
analysis of the model based on Markov chain Monte Carlo (MCMC). We
demonstrate efficacy of the approach using a simulation
study, and we present an application in which bird density is
estimated from standard point count data.

Our paper challenges two common assertions
in statistical ecology: first is that sample units should be
structured so as to ensure independence of the observable random
variable 
and second is that individual
information is needed to obtain
estimates of population size and density.
Our proposed model directly
refutes both claims and suggests new classes of
sampling designs and statistical models for making inferences about
animal demographic parameters.

\section{Sampling design and data}
\label{sdata}

We consider a sampling design in which animals are counted at $J$
traps on $K$ occasions. Although we use the term ``trap,'' anything
capable of recording counts of unmarked individuals could be used,
such as a camera or a human observer conducting a point count survey. The
sample occasion can be an arbitrary time
period, such as a single day in a camera trap study, or a 10 min
survey interval. Trap locations are characterized by the
coordinates, $\mathbf{X} = \{\mathbf{x}_j\} \in\mathbb{R}^2$ 
for $j = 1, 2,\ldots, J$.
The data resulting from this design are the $J \times K$
matrix of counts, $\mathbf{n}=\{n_{jk}\}; k = 1, 2,\ldots, K$.

Unlike similar count-based sampling protocols, this design aims to
induce correlation in the
neighboring counts by organizing the traps
sufficiently close together so that individual animals might be encountered
at multiple locations. Thus, we do not make the customary
assumptions that counts can be viewed as i.i.d. outcomes and
that no movement occurs between sampling occasions.
In the following sections we develop a
hierarchical model that describes the process by which such
correlation is induced and, by fitting this model to data,
we obtain estimates of density and related parameters.

\section{The hierarchical model}
\label{smodel}

\subsection{Data model}
\label{ssobs}

The data consist of the trap-specific counts ${\mathbf n}$ and the trap
coordinates $\mathbf{X}$.
The count data can be viewed as reduced information summaries of the
data that would be observed if all individuals in the population were
marked or otherwise distinguishable. Let $z_{ijk}$ represent the
encounter data for individual $i=1,\ldots,N$ at trap $j$ on occasion
$k$. If an individual can be detected at most once during a sampling
occasion, $z_{ijk}$ will be binary, or if individuals can be detected
multiple times during a single occasion, $z_{ijk}\ge0$ is an
integer. In standard capture--recapture studies,
$z_{ijk}$ is observed directly for captured individuals, and the
collection of observations for an individual is referred to as its
``encounter history.'' The encounter data are deterministically
related to the trap-level count data according to
\[
n_{jk} = \sum_{i=1}^{N}
z_{ijk}.
\]
However, we do not know $N$, and we cannot observe $z_{ijk}$
when individuals are unmarked.
Nonetheless, by developing our model in terms of these missing data, a
simplified analysis of the posterior is possible using classical data
augmentation methods. In particular, sampling the latent data
$\mathbf{z}_{jk} = \{z_{1jk}, z_{2jk}, \ldots, z_{Njk}\}$, conditional
on $n_{jk}$, uses an application of data augmentation
[\citet{tannerwong1987}] similar to that employed by
\citet{wolpertIck1998}. We will temporarily proceed by assuming that
$N$ is known so that we can focus on the detection process.

For the latent encounter data we assume
%
\begin{equation}\label{eqlatentPoisson}
z_{ijk} \sim\operatorname{Poisson}(\lambda_{ij}),
\end{equation}
where $\mathbb{E}(z_{ijk}) = \lambda_{ij}$ is the expected number of
captures or detections
of individual $i$ at trap $j$. We model this encounter rate
as a function of the Euclidean distance between the individual's
activity center $\mathbf{s}_i$ (also a latent variable) and the trap location,
$d_{ij} = \|\mathbf{s}_i - \mathbf{x}_j\|$. A
model for the activity centers is presented in the next section;
here we continue by assuming that the expected encounter frequency of an
individual is related to $d_{ij}$ as follows:
\[
\lambda_{ij} = \lambda_0 g(d_{ij}),
\]
where $\lambda_0$ is the encounter rate at $d=0$ and $g(d)$
is a positive-valued, typically monotonically decreasing,
function of distance. We make use of the standard half-normal detection
function used in distance sampling [\citet{bucklandetal2001}]:
\[
g(d) = \exp\bigl(-d^2 / 2\sigma^2\bigr),
\]
where $\sigma$ is a scale parameter determining the rate of decay in
encounter probability. This parameter also determines the degree of
correlation among
counts since animals with large home ranges are more likely to be
detected at multiple traps relative to animals with small home ranges.
This is analogous to correlation induced by averaging
spatial noise, in which case there is a well-defined relationship
between the
smoothing kernel and the induced covariance function
[\citet{higdon02}]. Finally, we note that although our focus here
is on distance-related
heterogeneity in encounter rate, other sources of
variation could be modeled by considering trap- or occasion-specific
covariates of
$\lambda_0$ and $\sigma$ as is customary in traditional SCR applications.

Under this formulation of the model based on data
augmentation---that is, including the latent encounter data in the
model---the full conditional of the latent encounter data is multinomial
\[
\{z_{1jk}, z_{2jk}, \ldots, z_{Njk}\} \sim
\operatorname{MN}\bigl(n_{jk}, \{\pi_{1j},
\pi_{2j},\ldots, \pi_{Nj}\}\bigr),
\]
%
where
$\pi_{ij} = \lambda_{ij} / \sum_i \lambda_{ij}$. A complete
description of all the full conditionals is provided
in the supplementary material [\citet{ChaRoy13}].

We note that the latent data model implies that the trap counts are
also Poisson:
%
\begin{equation}\label{eqnagg}
n_{jk} \sim\operatorname{Poisson}( \Lambda_{j} ),
\end{equation}
where
\[
\Lambda_{j} = \lambda_{0} \sum
_{i=1}^N g(d_{ij}),
\]
and the analysis can proceed from this model specification without
contemplating the
latent data.
Further, because $\Lambda_{j}$ does not depend on $k$, we can
aggregate the replicated counts, by a sufficiency argument, defining
$n_{j\cdot} = \sum_{k} n_{jk}$ and then
\[
n_{j\cdot} \sim\operatorname{Poisson}( K \Lambda_{j} ).
\]
As such, $K$ and $\lambda_{0}$ serve equivalent roles as affecting
baseline encounter rate.
This formulation of the model in terms of the aggregate counts
simplifies computations, as the unobserved encounter histories
do not need to be updated in the MCMC estimation
scheme. However, retaining the latent encounter data
in the formulation of the model is important if some individuals are
uniquely marked. In this case, modifying the MCMC algorithm to include
both types of data is trivial.

\subsection{Process model}
\label{ssstate}

The models for the data and the latent data are conditional on the
underlying ecological process of interest, which is the number and
locations of the unobserved activity centers $\{ \mathbf{s}_i \}$; $i=
1, 2,\ldots, N$.
We view the\vadjust{\goodbreak} activity centers as outcomes of
a spatial point process within a state-space, or observation window,
$\mathcal{S}$, which for simplicity we treat as planar $\cal{S} \subset
\mathbb{R}\mathrm{^2}$. Although any polygon
containing $\mathbf{X}$ could be considered, in
practice $\mathcal{S}$ should be chosen large enough so that
an individual's encounter rate is negligible if its activity center
occurs on the edge of the polygon. This will typically be a function
of the species' home range size. Alternatively, $\mathcal{S}$ may be
defined by geographic boundaries, such as when a species occurs on an
island; or it
may be defined based upon biological considerations such as suitable
habitat [\citet{royleEA09cameraEcol}].

In principle, general point process models could be considered
[\citet{borchersEfford08scr,illianetal2008}], but for simplicity we
adopt the homogeneous model 
\[
{\mathbf s}_{i} \sim\operatorname{Uniform}({\cal S}),
\]
which is to say that the point process intensity is constant $\mu
(x)_{x\in\mathcal{S}} =
N/\nu(\mathcal{S})$ where $\nu(\mathcal{S})$ is the area of the
state-space. Under this model, animals can move about their activity
centers, but the activity centers themselves do not move. Furthermore,
the activity centers are assumed to exhibit no
attraction or repulsion---assumptions that might not strictly hold
when animals exhibit behaviors such as territoriality. However, the
uniform model allows for \textit{any} realized configuration of
activity centers,
and, hence, the estimated locations of activity centers may
reflect departure from this assumption, albeit implicitly rather than
explicitly. 

Thus far we have treated $N$ as known, which implies that the model
for the activity centers is a binomial point process.
Although the model is naturally described conditional on $N$,
that is, in terms of $N$ latent encounter histories,
in all practical applications $N$ is unknown and, in fact, is the
object of
inference. 

\subsection{$N$ unknown}
\label{ssda}

The fact that $N$ is unknown presents a
technical challenge when implementing MCMC because the dimension
of the parameter space can change
with each Monte Carlo iteration, as the number of latent activity centers
changes. To resolve this, we expand our data augmentation scheme
following
\citet{royleEA07DA} and \citet{royledorazio2011} who proposed
fixing the
dimensions of parameter space by contemplating the existence of $M$,
rather than $N$ individuals in the population, where $M$ is some
integer $\gg N$. 
This strategy, known as parameter-expanded data augmentation
[\citet{liuwu1999}],
not only fixes the dimensions of
the problem, but it also allows for the specification of a discrete uniform
prior $N \sim\operatorname{DUnif}(0,M)$. We construct this prior by assuming
$N|M,\psi\sim\operatorname{Bin}(M,\phi)$ and $\phi\sim\operatorname{Unif}(0,1)$
which implies, marginally, that $N$ has the discrete uniform prior.
However,
the hierarchical formulation of the prior suggests an implementation
in which we introduce a set of latent indicator variables $w_{i} \sim
\operatorname{Bern}(\psi); i=1,2,\ldots,M$ and, furthermore, the model implies
that $z_{ijk}$ is obtained from the specified distribution
(\ref{eqlatentPoisson}) if $w_{i} = 1$. Alternatively, if
$w_{i}=0$, then $z_{ijk} =0$ with probability 1. In
effect, extending the model in this way induces a reparameterization
for the latent counts that is a zero-inflated version
of the original conditional-on-$N$ model. Specifically, the model
under parameter-expanded data augmentation becomes
\begin{eqnarray*}
z_{ijk}|w_{i} &\sim& \operatorname{Poisson}(\lambda_{ij}
w_i),
\\
w_{i} &\sim& \operatorname{Bern}(\psi)
\end{eqnarray*}
and, hence, $N = \sum_{i=1}^{M} w_i$ and population
density is simply $D = N/\nu(\mathcal{S})$. In general, $M$ should be large
enough such that the posterior of $N$ is unaffected by its choice,
that is, $\operatorname{Pr}(N=M) \approx0$; however, setting $M$ too high
will increase computation time unnecessarily.

\subsection{The joint posterior distribution}
\label{spost}

Assuming mutual independence of the hyperpriors,
that is, $[\psi,\lambda_0,\sigma] \propto[\psi][\lambda_0][\sigma]$, the
joint posterior distribution of the parameters is
%
\begin{eqnarray}\qquad
&&
[\mathbf{z}, \mathbf{w}, \mathbf{s}, \psi, \lambda_0, \sigma, |
\mathbf{n}, \mathbf{X}]
\nonumber\\[-8pt]\\[-8pt]
&&\qquad\propto\Biggl\{ \prod_{i=1}^M \Biggl\{ \prod
_{j=1}^J \prod
_{k=1}^K [n_{jk}|z_{ijk}]
[z_{ijk}|
w_i,\mathbf{s}_i,\sigma,
\lambda_0
] \Biggr\}[w_i|\psi] [
\mathbf{s}_i] \Biggr\}[\psi] [\lambda_0] [\sigma].
\nonumber
\end{eqnarray}
The only distributions not specified thus far
are $[\lambda_0]$ and $[\sigma]$, which should be
chosen to reflect prior knowledge or lack thereof. Examples
are presented in Section~\ref{sapp}.

We developed two distinct Metropolis-within-Gibbs MCMC algorithms for
this model [\citet{ChaRoy13}]. In the first, the missing encounter
data are sampled from their full conditionals,
which is useful when
one or more individual identities are available, in which case the
encounter frequencies $\mathbf{z}_{i}$ are observable for those
individuals. The second
formulation of the algorithm is unconditional on
the latent encounter frequencies. In that case, the marginal
distribution for
$n_{jk}$ is precisely equation (\ref{eqnagg}). This algorithm is more
computationally efficient because it avoids having to update the
missing $z_{ijk}$
of which there are many. A description of the two algorithms and the
full conditionals, along with \textbf{R} code to implement the
models, is presented in the supplementary material [\citet{ChaRoy13}].

\section{Applications}
\label{sapp}

\subsection{Simulation studies}
\label{sssim}

We carried out two simulation studies to evaluate the basic efficacy of
the estimator. In the first study, all individuals were unmarked and
we assessed posterior\vadjust{\goodbreak} properties by simulating replicate data sets
under varying degrees of correlation in the counts. In the second
study, we measured the improvements in posterior precision obtained by
marking a subset of the population.

To investigate the effects of correlation, we used a $15 \times15$
trap grid with unit spacing
and simulated scenarios with $\sigma\in\{0.5, 0.75, 1.0\}$.
We selected these values because $\sigma$ should not be too small
relative to the grid spacing or the counts are independent, that is, the
trap totals are then i.i.d. Poisson random variables. Similarly, $\sigma$
should not be too large relative to trap spacing or else again the
counts become i.i.d. Poisson random
variables. We note that trap spacing is widely recognized as being relevant
in the application of spatial capture--recapture models, where models
require observations of individuals at multiple traps, although to
this point in time little formal analysis of the design problem has been
done. For the other parameters in the model we considered $T\lambda_{0}
\in
\{2.5, 5.0\}$ and $N \in\{27$, $45$, $75$\} individuals distributed on
a $20
\times20$ unit state-space centered over the $15 \times15$ array of
trap locations.
This configuration implies a buffer of 3 units around the traps, which
was sufficiently large to ensure that encounter rate was negligible
for the values of $\sigma$ considered. We fit the model to 200
data sets for each of the 18 scenarios. For each simulation, we drew
32,000 posterior samples and discarded the initial 2000. We then
computed root-mean-squared-error (RMSE) for the posterior mean and
mode as well as coverage rates for the 95\% highest posterior density
(HPD) intervals.
Because our interest was in the performance
of the estimator in specific regions of the parameter space,
we emphasize that
our evaluation of the estimators is based on a \textit{frequentist evaluation}
of specific posterior features (mean or mode).
That is, we fixed the parameters and generated replicate
data sets under the specified model and then calculated RMSE by
averaging over the data-generating distribution
$(\mathrm{data}|\mathrm{parameters})$. Classical notions of
Bayesian optimality based on averaging over the posterior distribution
therefore do not apply.

Results of our first simulation study indicate that for the small level of
$\sigma$, the posterior mode, if regarded as a point estimator of $N$,
is approximately frequentist unbiased (Table \ref{tsim}). However, the
posterior distributions are skewed, which results in
posterior means exhibiting frequentist bias on the order of 5--10\%.
Substantial reductions in RMSE are realized
as effective encounter rate doubles
from 2.5 to~5.0 ($T=5$ to $T=10$). Coverage of 95\% HPD intervals is close
to nominal
for this case. Performance of the estimator deteriorates as the ratio
of $\sigma$ to trap spacing increases. For
$\sigma= 0.75$ the posterior distributions are centered approximately
over the data generating value (having nearly frequentist unbiased
modes), but the coverage is slightly lower than nominal as the
posterior becomes more strongly skewed. The general pattern holds for
the highest level of $\sigma= 1.0$.

\begin{table}
\tablewidth=290pt
\caption{Simulation results showing the bias and precision of the
posterior mean and mode for the population size parameter,
$N$. Proportion of 95\% highest posterior density intervals covering
the data generating value is also reported. $\lambda_0 = 0.5$ for
all cases}\label{tsim}
\begin{tabular*}{\tablewidth}{@{\extracolsep{\fill}}lld{2.0}d{2.1}d{2.2}d{2.1}d{2.2}c@{}}
\hline
$\bolds{\sigma}$ & $\bolds{N}$ & \multicolumn{1}{c}{$\bolds{T}$}
& \multicolumn{1}{c}{\textbf{Mean}} & \multicolumn{1}{c}{\textbf{RMSE}}
& \multicolumn{1}{c}{\textbf{Mode}} & \multicolumn{1}{c}{\textbf{RMSE}}
& \multicolumn{1}{c@{}}{\textbf{Coverage}} \\
\hline
0.50 & 27 & 5 & 30.0 & 8.12 & 26.9 & 6.77 & 0.965 \\
& & 10 & 28.9 & 5.39 & 27.3 & 5.14 & 0.970 \\
& 45 & 5 & 50.4 & 13.48 & 45.5 & 11.37 & 0.965 \\
& & 10 & 47.6 & 8.90 & 45.2 & 8.50 & 0.945 \\
& 75 & 5 & 83.2 & 19.94 & 75.3 & 16.92 & 0.945 \\
& & 10 & 78.7 & 13.54 & 74.6 & 12.77 & 0.945 \\
[6pt]
0.75 & 27 & 5 & 30.5 & 8.83 & 27.3 & 7.69 & 0.945 \\
& & 10 & 28.6 & 5.76 & 26.9 & 5.42 & 0.935 \\
& 45 & 5 & 52.6 & 15.63 & 46.6 & 13.95 & 0.950 \\
& & 10 & 49.5 & 11.38 & 45.9 & 10.91 & 0.925 \\
& 75 & 5 & 84.6 & 27.82 & 75.0 & 24.53 & 0.935 \\
& & 10 & 81.6 & 18.79 & 75.2 & 16.49 & 0.950 \\
[6pt]
1.00 & 27 & 5 & 32.7 & 12.90 & 28.0 & 11.06 & 0.920 \\
& & 10 & 30.0 & 7.72 & 27.4 & 6.87 & 0.925 \\
& 45 & 5 & 57.3 & 23.33 & 48.1 & 20.39 & 0.945 \\
& & 10 & 52.6 & 14.56 & 46.4 & 13.56 & 0.940 \\
& 75 & 5 & 90.3 & 36.55 & 76.0 & 38.10 & 0.930 \\
& & 10 & 87.1 & 26.62 & 75.9 & 25.83 & 0.975 \\
\hline
\end{tabular*}
\end{table}

To assess the influence of marking a subset of individuals, we used
the same number and configuration of traps as described above, and we
set $\sigma=0.5$, $\lambda_0=0.5$, $N=75$, and $T=5$. Then we generated
200 data sets
for $m \in\{5, 15, 25, 35\}$, where $m$ is the number of marked
individuals randomly sampled from the population.

\begin{figure}

\includegraphics{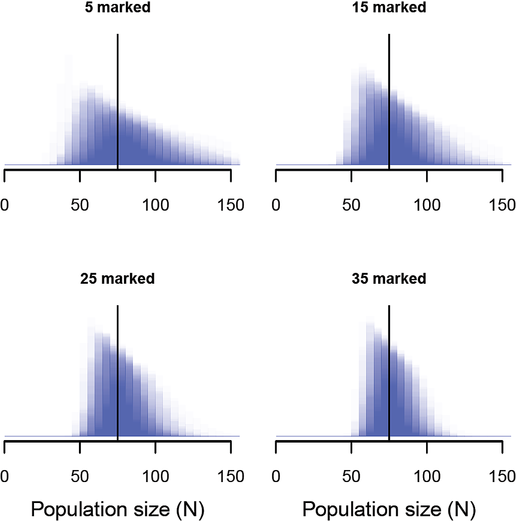}

\caption{Simulation results demonstrating the effect of marking
5, 15, 25, and 35 individuals
on the posterior distributions for population size. Each panel
shows 200 overlaid posterior distributions, represented as
histograms with transparent fill. The vertical line
is the data generating value $N=75$.}
\label{figNposts}
\end{figure}

Posterior distributions of $N$ for different numbers of marked
individuals are shown in Figure \ref{figNposts}. As anticipated,
posterior precision increases substantially with the
proportion of marked individuals. The posterior mode was
approximately unbiased as a point estimator, and RMSE
decreased 61\% from 17.31 when all 75~individuals were
unmarked to 6.82 when 35 individuals were marked
(Tables \ref{tsim} and \ref{tmarked}). Coverage was close to
nominal for all values of $m$ and posterior skew 
diminished as $m$ increased (Table \ref{tmarked}).

\begin{table}[b]
\tablewidth=270pt
\caption{Posterior mean, mode, and 95\% HPD interval coverage for
simulations in which $m$ of $N=75$ individuals were marked. Two
hundred simulations of each case were conducted}\label{tmarked}
\begin{tabular*}{\tablewidth}{@{\extracolsep{\fill}}lcd{2.2}cd{2.2}c@{}}
\hline
\textbf{\# Marked} & \multicolumn{1}{c}{\textbf{Mean}}
& \multicolumn{1}{c}{\textbf{RMSE}} & \multicolumn{1}{c}{\textbf{Mode}}
& \multicolumn{1}{c}{\textbf{RMSE}} & \multicolumn{1}{c@{}}{\textbf{Coverage}} \\
\hline
$m=5$ & 80.1 & 14.53 & 75.9 & 13.88 & 0.955 \\
$m=15$ & 78.4 & 11.59 & 75.9 & 11.26 & 0.945 \\
$m=25$ & 77.6 & 8.51 & 75.7 & 8.40 & 0.960 \\
$m=35$ & 77.0 & 6.92 & 75.3 & 6.82 & 0.960 \\
\hline
\end{tabular*}
\end{table}

\subsection{Point count data}
\label{ssptct}

\begin{figure}

\includegraphics{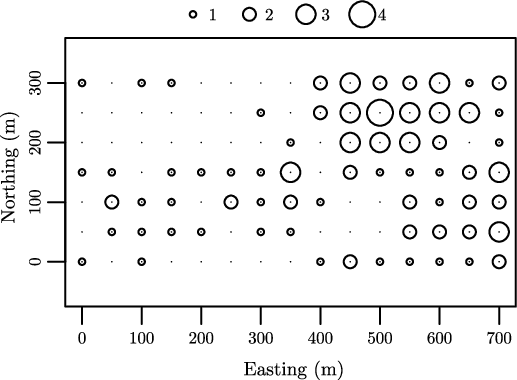}

\caption{Spatially correlated counts of northern parula on a 50 m
grid. The size of the circle represents the total number of
detections at each point.}
\label{fignopaDat}
\end{figure}

We applied our model to point
count data collected on the northern parula (\textit{Parula americana}), a
migratory passerine. This species maintains
well-defined home ranges during the breeding season
[\citet{MoldenhaerNOPAbna}], and thus our modeling effort was focused
on estimating the number and location of home range centers. Points
were located on a 50 m grid, which ensured spatial
correlation since home ranges typically have $>$50 m radii and because
their song can be heard from distance
$>$50 m [\citet{MoldenhaerNOPAbna}]. This small grid spacing contrasts
with the conventional
practice of spacing points by $>$200 m to obtain i.i.d.
counts. Figure \ref{fignopaDat} depicts the spatially correlated
counts ($n_{r\cdot}$) from the 105 point count locations
surveyed three times each during June 2006
at the Patuxent Wildlife Research Center in Laurel Maryland, USA.
A~total of 226 detections were made with a maximum count of 4 during a
single survey. At 38 points, no Parulas were detected. All but one of
the detections were of singing males, and this one observation was
not included in the analysis.

In our analysis of the Parula data, we defined the point process
state-space by buffering the grid of point
count locations by 250 m and used $M=300$ for data augmentation. We
simulated posterior
distributions using three Markov chains,
each consisting of 300,000 iterations after discarding the initial 10,000
draws. Convergence was indicated by visual inspections of the Markov
chain histories, and by $\hat{R}$ statistics [\citet{gelmanRubin92}]
$<$1.1 for each of the monitored parameters: $\lambda_0$,
$\sigma$, and $N$. The history plots are available in the supplementary
material [\citet{ChaRoy13}].

\begin{table}
\caption{Posterior summary statistics for the spatial
model applied to the northern parula data. Two sets of priors for
the encounter rate parameter $\sigma$ were considered. $M=300$ was
used in both cases. Parulas/ha, $D$, is a derived parameter}\label{tnopaPosts}
\begin{tabular*}{\tablewidth}{@{\extracolsep{\fill}}l c d{2.2}d{2.2}d{2.2}d{2.3}d{2.2}d{3.2}@{}}
\hline
\textbf{Par} & \multicolumn{1}{c}{\textbf{Prior}} & \multicolumn{1}{c}{\textbf{Mean}}
& \multicolumn{1}{c}{\textbf{SD}} & \multicolumn{1}{c}{\textbf{Mode}}
& \multicolumn{1}{c}{\textbf{q0.025}} & \multicolumn{1}{c}{\textbf{q0.50}}
& \multicolumn{1}{c@{}}{\textbf{q0.975}} \\
\hline
$\sigma$ & $U(0, \infty)$ & 2.15 & 1.22 & 1.23 & 0.90 & 1.67 & 5.17 \\
$\lambda_0$ & $U(0, \infty)$ & 0.28 & 0.15 & 0.21 & 0.08 & 0.26 & 0.67
\\
$N$ & $U(0, M)$ & 40.95 & 38.07 & 4.00 & 3.00 & 31.00 & 143.00 \\
$D$ & -- & 0.43 & 0.40 & 0.04 & 0.031 & 0.32 & 1.49 \\
[6pt]
$\sigma$ & $G(13, 10)$ & 1.30 & 0.26 & 1.23 & 0.90 & 1.27 & 1.91 \\
$\lambda_0$ & $U(0, \infty)$ & 0.30 & 0.13 & 0.24 & 0.10 & 0.28 & 0.60
\\
$N$ & $U(0, M)$ & 59.32 & 36.49 & 36.00 & 18.00 & 50.00 & 157.00 \\
$D$ & -- & 0.62 & 0.38 & 0.38 & 0.19 & 0.52 & 1.64 \\
\hline
\end{tabular*}
\end{table}

One benefit of a Bayesian analysis is that it can accommodate prior
information about home range size, 
which is readily available for many
species and directly related to the encounter rate parameter $\sigma$
[\citet{royleEA11search}]. To illustrate, we analyzed the Parula
data using two sets of
priors. In the first set, all priors were
improper, customary noninformative priors (see Table \ref{tnopaPosts}).
Uniform priors were also used in the second set, with the exception of
an informative prior for the scale parameter $\sigma\sim
\operatorname{Gamma}(13,10)$. We arrived at this prior using the methods
described by \citet{royleEA11search} and published
information on home range size and detection probability
[\citet{MoldenhaerNOPAbna,SimonsEA09}]. More details on this
derivation are found in the supplementary material [\citet
{ChaRoy13}]. We briefly note here that this prior
includes the biologically plausible range of values for $\sigma$
suggested by the published literature.

\begin{figure}[b]

\includegraphics{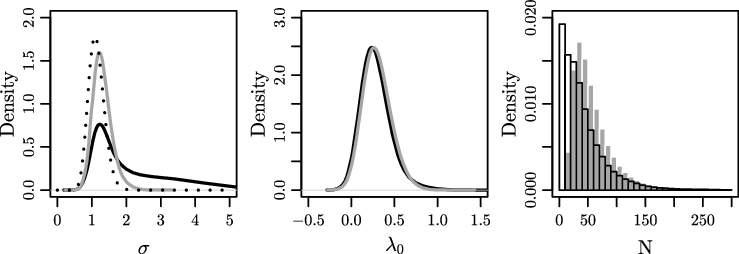}

\caption{Effects of $\sigma\sim\operatorname{Gamma}(13,10)$
prior on the posterior distributions from the northern parula
model. Posteriors from the model with uniform priors are
shown in black, and posteriors from the informative prior model
are shown in gray. The prior itself is shown as a dotted line in the
leftmost panel.}
\label{figprior}
\end{figure}

The posterior distribution for
$N$ was highly skewed with a long right tail resulting in a wide 95\%
credible interval (Table \ref{tnopaPosts}). Nonetheless, the interval
for density,~$D$, includes estimates reported from more intensive field
studies [\citet{MoldenhaerNOPAbna}]. This was true when considering
both sets of priors, although posterior precision was higher under the
informative set of priors. Specifically, the use of prior information
reduced posterior density at high, biologically implausible,
values of $\sigma$, and hence decreased the posterior mass for
low values of $N$ (Figure \ref{figprior}). For both sets of priors,
$\operatorname{Pr}(N=M=300) \approx0$, indicating that the amount of data
augmentation was sufficient to avoid any effect on the posteriors.

\begin{figure}

\includegraphics{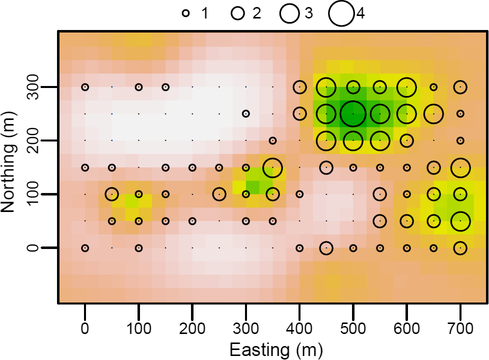}

\caption{Estimated density surface of northern parula activity
centers. The grid of point count locations with count totals is
superimposed. See Figure \protect\ref{figNposts} for additional details.}
\label{fignopaDen}
\end{figure}

In addition to estimating density, our model can be used to produce
density surface maps, which are often used in applied ecological
research to direct management efforts and develop hypotheses regarding
the factors influencing abundance.
Density surface maps can be produced by discretizing the
state-space and tallying the number of activity centers occurring in
each pixel during each MCMC iteration. Parula density was
highest near the northeastern corner of the study plot, which may
correspond to important habitat features such as suitable nest site
locations (Figure \ref{fignopaDen}). We anticipate future model
extensions to directly model the
point process intensity using habitat covariates.

\section{Discussion}
\label{sdis}

In this paper we confronted one of the most difficult challenges
faced in wildlife sampling---estimation of population density in the
absence of
data to distinguish among individuals. To do so, we developed a novel
class of spatially explicit models that
applies to spatially organized counts, where the count locations or
traps are located sufficiently close together so that individuals
are exposed to encounter at multiple traps. This design yields
correlation in the observed counts, and this correlation proves to be
informative about encounter rate parameters and density.
We note that sample locations in count based studies are typically
not organized close
together in space because conventional wisdom and standard practice
dictate that independence of sample units is necessary
[\citet{hurlbert84pseudorep}]. Our model
suggests that in some cases it might be advantageous to deviate from
the conventional wisdom if one is interested in inference about
density. Of course, this is also known in the application of standard spatial
capture--recapture models [\citet{borchersEfford08scr}]
where individual
identity is preserved across trap encounters, but it is seldom, if
ever, considered in the design of more traditional count surveys.

Our model has broad relevance to a vast number of animal
sampling problems. The motivating problem involved bird point counts
where individual
identity is typically not available. The model also applies
to other standard methods used to sample unmarked
populations, such as camera traps
or even methods that yield sign (e.g., scat, track) counts
indexed by space. However, results of our simulation study reveal some
important limitations of the basic
estimator applied to situations in which none of the individuals can
be uniquely identified. In particular, posterior
distributions are highly skewed in typical small to moderate sample
size situations and posterior precision is low,
although for more
expansive trapping grids, better performance can be expected.

Several modifications of the model can lead to improved
performance of the estimator.
Our simulation results demonstrate that marking a subset of
individuals can yield substantial increases in posterior
precision. Marking a subset of individuals is
commonplace in animal studies such as when a small number of
individuals are
radio-collared in conjunction with a count based survey
[\citet{bartmannEA87}]. In many other situations a subset of
individuals can be identified by natural marks alone
[\citet{kellyetal2008}] and, thus, our
model could be applied to data from camera trapping studies of
species such as mountain lions, deer, or coyotes.
The ability to study partially marked populations
adds flexibility to existing SCR methods and also
creates new opportunities for designing efficient SCR studies
since the costs of marking all individuals in a population can be
prohibitive.

When including data from marked individuals, it is
important to note that we assume that the marks can be reliably read
in the field, that is, there is no misidentification of marked
animals. If some marked individuals cannot be reliably recognized,
perhaps due to blurry photographs in camera trapping studies or DNA
samples that do not amplify, then the
questionable data records should be discarded so as not to bias
estimators. Explicitly
modeling misidentification of marked individuals deserves additional
study.

When applied to data from marked and unmarked individuals, our model
can be viewed as a spatial extension of traditional
``mark-resight'' estimators
[\citet{bartmannEA87,mintaMangel89,mcclintockHoeting09}]. In their
simplest form, mark-resight methods involve fitting standard
closed population capture--recapture models to the data on marked
individuals, and the resultant estimate of detection probability
($\hat{p}$) is used to estimate population size as $\hat{N} = m +
u/\hat{p}$, where $m$ and $u$ are the number of
marked and unmarked individuals, respectively. In addition to the
problem of converting $N$ to density,
the unmarked individuals provide no information about the
encounter rate parameters, and thus mark-resight methods cannot be
used unless a large sample of marked individuals is available. This
contrasts with our approach which can be used even when all
individuals are unmarked.

In some cases, such as in point counts of birds, it may not be
practical to mark individuals. An alternative to increasing posterior
precision is to utilize prior information on
home range size. Indeed, extensive information on home range size has
been compiled for many species in diverse habitats 
[e.g., \citet{degraafYamasaki01}]. It is
easy to embody this information in a prior distribution as we
demonstrated for the Parula data.

An additional design extension that could increase precision is to use
multiple sampling methods [\citet{gopalaswamyetal2012}], in which
one method generates encounter
frequencies and the other method generates individuality.
For example, camera traps are now commonly used with surveys for
sign (scat or tracks) or hair snares for sampling bear populations.
These distinct methods would have different basal detection
rates but share an underlying spatial model describing the
organization of individuals in space.
Our model shows promise for using
these disparate data types efficiently
for estimating density.

\subsection{$N$-mixture models}

Parallel developments which appear ostensibly unrelated to SCR models
have addressed the problem of estimating population size when
individuals are unmarked. $N$-mixture models [Royle
(\citeyear{royle04a,royle04b}), \citet{royledistsamp2004}] can be
applied to a repeated measures type of data structure wherein data are
collected at $J$ sites, with $K$ replicate surveys conducted at each.
$N$-mixture models regard abundance at each site ($N_j$) as an i.i.d.
realization from a discrete distribution, such as the Poisson or
negative binomial with expectation $\theta$. In the standard binomial
$N$-mixture model, the observed counts are treated as binomial outcomes
with $N_j$ ``trials'' and detection probability $p$.

Although these models have proven useful for studies of factors that
affect variation in abundance, interpretation of model parameters is
strongly dependent on the assumption that populations are
closed with respect to demographic processes and movement. The closure
assumption can be an important practical limitation
[but see \citet{dailMadsen11,chandlerEA11tempem}]. Furthermore, the
i.i.d. assumption is violated if spatial
correlation exists among sites, such as if animals move among plots.
Although we formulated the model developed in our paper as
an extension of SCR models, it clearly can
also be viewed as a spatially explicit extension of $N$-mixture models
where the local population sizes $N_j$ are dependent owing to the
nature of the sampling design.

Thus, two recently developed methodological frameworks, spatial
capture--recapture and $N$-mixture models, address
different problems that arise in sampling animal populations. SCR
models address nonclosure by accommodating information on
the juxtaposition of animal activity centers and traps, and
$N$-mixture models address the inability to uniquely identify individuals.
Our model unifies these two modeling frameworks by
addressing both issues simultaneously.

\subsection{Alternative observation models}
\label{ssext}

Several aspects of our 
model can be modified to accommodate
alternative sampling designs or parametric distributions.
We considered situations in which an individual can be detected more than
once at a trap during a single occasion, but under some designs this
is not possible. When collecting DNA samples, for instance, an
individual can often be detected at most once during an
occasion, because multiple samples of biological material cannot be
attributed to distinct episodes. Therefore, rather than $z_{ijk} \sim
\operatorname{Pois}(\lambda_{ij})$, we have $z_{ijk} \sim\operatorname
{Bern}(p_{ij})$
where, for example, $p_{ij} = p_0 \exp(-d_{ij}^2/(2\sigma^2))$, and
$p_0$ is the probability of detecting an individual whose home range
is centered on trap~$j$. This Bernoulli model is a focus of ongoing
investigations.

Both the Poisson and the Bernoulli models
produce count observations that when aggregated over individuals form
trap-specific totals; however, ecologists often collect
``detection/nondetection'' data because it can be easier to determine
if $\ge$1 individual was present rather than enumerating all
individuals in a location. In this case, the underlying $z_{ijk}$
array is the same as the above cases, but we observe $n_{jk} =
I(\sum_{i=1}^{N} z_{ijk} > 0)$ where $I$ is the indicator
function. This model is
a spatially explicit extension of the model of
\citet{royleNichols03} in which the underlying abundance state
is inferred from binary data. We have investigated this model to a
limited extent but do not report on those results here.

\subsection{Spatial point process models}
\label{sssimilar}

Our model has some direct linkages to existing point process
models. We note that the observation intensity function (i.e.,
corresponding to the observation
locations) is a compound Gaussian kernel similar to
that of the Thomas process
[\citet{mollerwaagepetersen2003}, pages 61 and 62,
\citet{thomas1949}].
Also, the Poisson-Gamma convolution models
[\citet{wolpertIck1998}] are structurally similar [see also
\citet{higdon1998process}
and \citet{best2000spatial}]. In particular, our model is such a
model but
with a \textit{constant} basal encounter rate $\lambda_{0}$
and \textit{unknown} number and location of ``support points,'' which in
our case are the animal activity centers, $\{\mathbf{s}_i\}$.
We can thus regard our model as an approach for
estimating the location and local density of support points,
which we believe could be useful in the application of
convolution models. \citet{best2000spatial} devise an MCMC
algorithm for the
Poisson-Gamma model based on data augmentation, which is
similar to the component of our algorithm for updating the missing data
in the conditional-on-$\mathbf{z}$ formulation of the model. We
emphasize that
our model is distinct from these Poisson-Gamma models
in that we estimate the number \textit{and} location of such
support points.

If individuals were perfectly observable, then the resulting point
process of locations is clearly a standard Poisson or binomial (fixed
$N$) cluster process or Neyman--Scott process.\vadjust{\goodbreak}
If detection is uniform over space but
imperfect, then the basic process is unaffected by this random thinning.
Our model can therefore be viewed formally as a Poisson (or binomial)
cluster process model, but one in which the thinning is
nonuniform, governed by the encounter model which dictates that the
thinning rate increases with distance from the observation points. In
addition, our inference objective is, essentially, to estimate the
number of parents in the underlying Poisson cluster
process, where the observations are biased by an incomplete sampling
apparatus (points in space).

As a model of a thinned point process, our model has much in common
with classical distance sampling models
[\citet{bucklandetal2001,johnsonetal2009}].
The main distinction is that our data structure does not include
observed distances, although the underlying observation model is
fundamentally the same as in distance sampling if there is only a
single replicate sample and $\mathbf{s}_i$ is defined as an individual's
location at an instant in time. For replicate samples, our model preserves
(latent) individuality across samples and traps which is not a feature
of distance sampling. We note that error in measurement of distance is
not a relevant consideration in our model, and we do not
require the standard distance sampling assumption that the probability
of detection is 1 if an individual occurs at the survey point. More
importantly, distance sampling models cannot be applied to data from
many of the sampling designs for which our model is relevant. For
example, many rare and endangered species can only be
effectively surveyed using noninvasive methods such as hair snares and camera
traps that do not produce distance data [\citet{oconnell10book}].

\section{Conclusion}

Concerns about statistical independence have prompted
ecologists to design count based studies such that the observed
random variables can be regarded as i.i.d. outcomes
[\citet{hurlbert84pseudorep}]. Interestingly, this
often proves impossible in practice, and elaborate
methods have been devised to model spatial dependence as a nuisance
parameter. Conversely,
our view is that spatial dependence is an important element of the
underlying ecological process which is of direct interest in
ecological investigations.
Our paper presents a modeling framework that directly
confronts the classical view of spatial dependence as a nuisance
by demonstrating that spatial
correlation carries information about the locations of individuals,
which can be used to estimate density even when individuals
are unmarked and distance-related heterogeneity exists in encounter
probability.

\section*{Acknowledgments}

We thank the Associate Editor, two anonymous referees, R. Dorazio, and
A. O'Connell for many helpful suggestions. The U.S. FWS Black Duck
Joint Venture provided funding to J. A. Royle for some early work on
this problem. We thank D. Dawson, USGS, for collecting the point count
data.

\begin{supplement}
\stitle{Full conditional distributions, \textbf{R} code, and history plots\\}
\slink[doi]{10.1214/12-AOAS610SUPP} 
\sdatatype{.zip}
\sfilename{aoas610\_supp.zip}
\sdescription{Supplement A1 is a description of the full conditional
distributions. Supplement A2 includes \textbf{R} code for implementing
the MCMC
algorithms and simulating data. It also contains the northern
parula data set and a \mbox{description} of the method used to obtain the
informative prior used in the analysis of the Parula
data. Supplement A3 is a panel of history plots for the Markov
chains from the northern parula analysis.}
\end{supplement}

%

\printaddresses

\end{document}